\documentclass[twocolumn,prd,superscriptaddress,altaffilletter,nofootinbib]{revtex4-1}

\usepackage{lineno}
\usepackage{amsmath}
\usepackage{amssymb}
\usepackage{amsfonts}
\usepackage{graphicx}
\usepackage{hyperref}
\usepackage[usenames,dvipsnames]{xcolor}
\usepackage{xspace}

\allowdisplaybreaks

\allowdisplaybreaks

\newcommand{\data}{\ensuremath{\mathbf{s}}}
\newcommand{\signal}{\ensuremath{\mathbf{h}}}
\newcommand{\noise}{\ensuremath{\mathbf{n}}}

\newcommand{\g}{\ensuremath{\mathbf{g}}}

\newcommand{\M}{\ensuremath{\mathbf{M}}}
\newcommand{\I}{\ensuremath{\mathbf{I}}}
\newcommand{\params}{\ensuremath{\vec{\theta}}}
\newcommand{\inparams}{\ensuremath{\vec{\lambda}}}
\newcommand{\x}{\inparams}
\newcommand{\xs}{\params}
\newcommand{\dx}{\ensuremath{\vec{\Delta\lambda}}}
\newcommand{\dxs}{\ensuremath{\Delta\xs}}
\newcommand{\fft}[1]{\ensuremath{\mathcal{F}\left[#1\right]}}
\newcommand{\ifft}[1]{\ensuremath{\mathcal{F}^{-1}\left[#1\right]}}

\begin{document}

\title{Metric Assisted Stochastic Sampling (MASS) search for gravitational waves from binary black hole mergers}

\author{Chad Hanna}
\affiliation{Department of Physics, The Pennsylvania State University, University Park, PA 16802, USA}
\affiliation{Institute for Gravitation and the Cosmos, The Pennsylvania State University, University Park, PA 16802, USA}
\affiliation{Department of Astronomy and Astrophysics, The Pennsylvania State University, University Park, PA 16802, USA}
\affiliation{Institute for Computational and Data Sciences, The Pennsylvania State University, University Park, PA 16802, USA}

\author{Prathamesh Joshi}
\email[]{ppj5075@psu.edu}
\affiliation{Department of Physics, The Pennsylvania State University, University Park, PA 16802, USA}
\affiliation{Institute for Gravitation and the Cosmos, The Pennsylvania State University, University Park, PA 16802, USA}

\author{Rachael Huxford}
\affiliation{Department of Physics, The Pennsylvania State University, University Park, PA 16802, USA}
\affiliation{Institute for Gravitation and the Cosmos, The Pennsylvania State University, University Park, PA 16802, USA}

\author{Kipp Cannon}
\affiliation{RESCEU, The University of Tokyo, Tokyo, 113-0033, Japan}

\author{Sarah Caudill}
\affiliation{Nikhef, Science Park, 1098 XG Amsterdam, Netherlands}

\author{Chiwai Chan}
\affiliation{RESCEU, The University of Tokyo, Tokyo, 113-0033, Japan}

\author{Bryce Cousins}
\affiliation{Department of Physics, The Pennsylvania State University, University Park, PA 16802, USA}
\affiliation{Institute for Computational and Data Sciences, The Pennsylvania State University, University Park, PA 16802, USA}

\author{Jolien D. E. Creighton}
\affiliation{Leonard E.\ Parker Center for Gravitation, Cosmology, and Astrophysics, University of Wisconsin-Milwaukee, Milwaukee, WI 53201, USA}

\author{Becca Ewing}
\affiliation{Department of Physics, The Pennsylvania State University, University Park, PA 16802, USA}
\affiliation{Institute for Gravitation and the Cosmos, The Pennsylvania State University, University Park, PA 16802, USA}

\author{Miguel Fernandez}
\affiliation{Department of Physics, The Pennsylvania State University, University Park, PA 16802, USA}
\affiliation{Institute for Gravitation and the Cosmos, The Pennsylvania State University, University Park, PA 16802, USA}

\author{Heather Fong}
\affiliation{RESCEU, The University of Tokyo, Tokyo, 113-0033, Japan}
\affiliation{Graduate School of Science, The University of Tokyo, Tokyo 113-0033, Japan}

\author{Patrick Godwin}
\affiliation{Department of Physics, The Pennsylvania State University, University Park, PA 16802, USA}
\affiliation{Institute for Gravitation and the Cosmos, The Pennsylvania State University, University Park, PA 16802, USA}

\author{Ryan Magee}
\affiliation{LIGO Laboratory, California Institute of Technology, MS 100-36, Pasadena, California 91125, USA}

\author{Duncan Meacher}
\affiliation{Leonard E.\ Parker Center for Gravitation, Cosmology, and Astrophysics, University of Wisconsin-Milwaukee, Milwaukee, WI 53201, USA}

\author{Cody Messick}
\affiliation{LIGO Laboratory, Massachusetts Institute of Technology, Cambridge, MA 02139, USA}

\author{Soichiro Morisaki}
\affiliation{Institute for Cosmic Ray Research, The University of Tokyo, 5-1-5 Kashiwanoha, Kashiwa, Chiba 277-8582, Japan}

\author{Debnandini Mukherjee}
\affiliation{Department of Physics, The Pennsylvania State University, University Park, PA 16802, USA}
\affiliation{Institute for Gravitation and the Cosmos, The Pennsylvania State University, University Park, PA 16802, USA}

\author{Hiroaki Ohta}
\affiliation{RESCEU, The University of Tokyo, Tokyo, 113-0033, Japan}

\author{Alexander Pace}
\affiliation{Department of Physics, The Pennsylvania State University, University Park, PA 16802, USA}
\affiliation{Institute for Gravitation and the Cosmos, The Pennsylvania State University, University Park, PA 16802, USA}

\author{Stephen Privitera}
\affiliation{Albert-Einstein-Institut, Max-Planck-Institut für Gravitationsphysik, D-14476 Potsdam-Golm, Germany}

\author{Surabhi Sachdev}
\affiliation{Department of Physics, The Pennsylvania State University, University Park, PA 16802, USA}
\affiliation{Institute for Gravitation and the Cosmos, The Pennsylvania State University, University Park, PA 16802, USA}
\affiliation{LIGO Laboratory, California Institute of Technology, MS 100-36, Pasadena, California 91125, USA}

\author{Shio Sakon}
\affiliation{Department of Physics, The Pennsylvania State University, University Park, PA 16802, USA}
\affiliation{Institute for Gravitation and the Cosmos, The Pennsylvania State University, University Park, PA 16802, USA}

\author{Divya Singh}
\affiliation{Department of Physics, The Pennsylvania State University, University Park, PA 16802, USA}
\affiliation{Institute for Gravitation and the Cosmos, The Pennsylvania State University, University Park, PA 16802, USA}

\author{Ron Tapia}
\affiliation{Department of Physics, The Pennsylvania State University, University Park, PA 16802, USA}
\affiliation{Institute for Computational and Data Sciences, The Pennsylvania State University, University Park, PA 16802, USA}

\author{Leo Tsukada}
\affiliation{RESCEU, The University of Tokyo, Tokyo, 113-0033, Japan}
\affiliation{Graduate School of Science, The University of Tokyo, Tokyo 113-0033, Japan}

\author{Daichi Tsuna}
\affiliation{RESCEU, The University of Tokyo, Tokyo, 113-0033, Japan}

\author{Takuya Tsutsui}
\affiliation{RESCEU, The University of Tokyo, Tokyo, 113-0033, Japan}

\author{Koh Ueno}
\affiliation{RESCEU, The University of Tokyo, Tokyo, 113-0033, Japan}

\author{Aaron Viets}
\affiliation{Leonard E.\ Parker Center for Gravitation, Cosmology, and Astrophysics, University of Wisconsin-Milwaukee, Milwaukee, WI 53201, USA}

\author{Leslie Wade}
\affiliation{Department of Physics, Hayes Hall, Kenyon College, Gambier, Ohio 43022, USA}

\author{Madeline Wade}
\affiliation{Department of Physics, Hayes Hall, Kenyon College, Gambier, Ohio 43022, USA}

\author{Jonathan Wang}
\affiliation{Department of Physics, University of Michigan, Ann Arbor, Michigan 48109, USA}

\date{\today}

\begin{abstract}
We present a novel gravitational wave detection algorithm that conducts a
matched filter search stochastically across the compact binary parameter space
rather than relying on a fixed bank of template waveforms.  This technique is
competitive with standard template-bank-driven pipelines in both 
computational cost and sensitivity.  However, the complexity of the analysis is
simpler allowing for easy configuration and horizontal scaling across
heterogeneous grids of computers.  To demonstrate the method we analyze
approximately one month of public LIGO data from July 27 00:00 2017 UTC -- Aug
25 22:00 2017 UTC and recover eight known confident gravitational wave
candidates.  We also inject simulated binary black hole (BBH) signals to
demonstrate the sensitivity.
\end{abstract}

\maketitle

\section{Introduction}

Advanced LIGO directly detected gravitational waves (GWs) for the first time in
2015 from the merger of two black holes each about 30 times the mass of our
Sun~\cite{Abbott:2016blz}.  The second confident binary black hole (BBH)
observation came just three months later~\cite{Abbott:2016nmj}. Since then, the
LIGO and Virgo Collaborations have detected a total of 90 compact binary
mergers~\cite{LIGOScientific:2018mvr, abbott2020gwtc, abbott2021gwtc, theligoscientificcollaboration2021gwtc3}, including two neutron
star mergers~\cite{TheLIGOScientific:2017qsa, Abbott:2020uma} and two neutron star-black hole mergers~\cite{abbott2021observation}.  LIGO and Virgo
have made their data public~\cite{Trovato:2019liz} resulting in several new BBH
discoveries by the community~\cite{Zackay:2019btq, Venumadhav:2019lyq,
Zackay:2019tzo, Venumadhav:2019tad, nitz20213ogc, Abbott_2021}.

Historically, gravitational wave searches for compact binary coalescence have
relied on matched filtering~\cite{Finn:1992xs, PhysRevD.53.6749, Owen:1998dk},
with several groups building on matched filtering as the foundation for their
algorithms~\cite{Allen:2005fk, Cannon:2011vi, babak2013searching, Nitz:2017svb,
Adams:2015ulm, Venumadhav:2019tad}.  These techniques rely on fixed
banks of templates~\cite{PhysRevD.53.6749, harry2009stochastic,
ajith2014effectual} and are known to scale poorly to high dimensional
spaces~\cite{harry2016searching}.  Stochastic sampling methods were first
proposed to address gravitational wave detection in future searches for
gravitational waves with LISA~\cite{cornish2005lisa}, but have not been widely
used for detection in LIGO and Virgo data.  Stochastic sampling techniques are,
however, state-of-the art for the estimation of compact binary parameters once
detections have been made~\cite{Veitch:2014wba, ashton2019bilby}.

In this work we blend aspects of traditional matched filter searches, bank
placement techniques, and stochastic sampling to create a new bank-less matched
filter search for gravitational waves. While it remains to be seen what the
broad applications of these techniques could be, we demonstrate a useful case
study here by analyzing LIGO data from the Hanford and Livingston detectors 
from August 2017~\cite{RICHABBOTT2021100658} to search for binary black
hole mergers.  We recover eight known gravitational wave candidates.

\section{Motivation}
Our goal is to develop an \textit{offline} compact binary search pipeline which
is designed to detect gravitational waves in archival, LIGO, Virgo, and KAGRA
data based on the GstLAL framework~\cite{Messick:2016aqy, Sachdev:2019vvd,
cannon2021gstlal, gstlal}.  We distinguish that an offline analysis has less
strict time-to-solution requirements (hours or days) compared to low-latency
analysis where the time-to-solution needs to be seconds.  We will not strive to
reach the time-to-solution needs of low-latency analysis with the algorithm we
present here.  Our motivation for revisiting offline matched filter detection
for gravitational waves is to more easily parallelize and deploy analysis
across heterogeneous resources such as multiple concurrent sites on the LIGO
and Virgo data grids, the Open Science Grid, campus resources, and commercial
clouds. We aim to achieve this by having a simpler workflow than competing
pipelines such as GstLAL. We also wish to simplify the setup required to conduct an analysis and
to improve usability for new researchers wanting to learn about gravitational
wave detection at scale.  The intersection of these desires led us to consider
new algorithmic approaches to searching the compact binary parameter space.

The Open Science Grid defines criteria for opportunistic computing as an
application that ``does not require message passing...has a small runtime
between 1 and 24 hours...can handle being unexpectedly killed and restarted..."
and ``...requires running a very large number of small jobs rather than a few
large jobs."~\cite{osgreqs}. Our proposed workflow consists of parallel jobs
that each search a small amount of gravitational-wave data from LIGO, Virgo and
KAGRA without any interdependency between jobs. To contrast, the current GstLAL
analysis workflow consists of a directed acyclic graph with more than ten
levels of interdependent jobs. In this new approach, we target a $\sim$1--12
hour runtime for each job, the use of one CPU core per job, and $\sim 2$ GB of
RAM required per job in order to maximize throughput on opportunistic compute
resources.  Each job implements a flexible checkpointing procedure allowing 
work to be periodically saved.

\section{Methods}
In this work, we will conduct a matched-filter search for binary black holes
with the goal of identifying the maximum likelihood parameters for candidate
events over 4s coalescence-time windows using an analysis that foregoes the
use of a pre-computed template bank and instead employs stochastic sampling of
the binary parameter space.  Our workflow consists of two stages. The first
stage executes N parallel jobs that conduct the bulk of the cpu-intensive work - 
in this study, this first stage consisted of 2974 such jobs.
The results of these parallel jobs are returned to a single location at which
point a second stage is run to combine results, assess candidate significance,
estimate the search sensivity and visualize the results.  This second stage
requires significantly lower computing power than the first stage, but is I/O
intensive and is designed to be run potentially on local resources after grid
jobs have completed.
  
In stage one, we begin by reading in gravitational
wave data from each observatory. Next, we measure the data noise power spectrum
and whiten the data using the inferred spectrum.  We then stochastically sample
the data by proposing jumps governed by a parameter space metric as described
in Section III-D. For each jump, we generate the appropriate template waveform
and then compute the matched filter signal-to-noise ratio (SNR) over a 6s 
stretch of time using 122s of data per calculation. 

Within a 4s time window, we identify peaks
in the matched filter output, known as triggers, for each detector that is
being analyzed. For each collection of triggers, we perform signal consistency
checks~\cite{Messick:2016aqy}, and calculate a likelihood ratio ranking
statistic~\cite{Cannon:2015gha}.  If the new sample has a larger SNR
than the previous sample, it is stored - otherwise a new jump is
proposed.

A local estimate of the noise background is obtained by forming
synthetic events from disjoint windows. This causes the time and phase difference between
detectors of a single background event to be uniformly distributed, which is what we expect
from noise events. This is a somewhat hybrid approach
between the time-slide method~\cite{babak2013searching} and sampling
methods~\cite{Cannon:2012zt} already employed in GW searches.  The second stage
gathers the candidate events, results of the simulated GW search, and the
background samples to produce a final summary view of the analysis results. In order to
estimate the sensitivity of our methods to detecting gravitational waves, we
conduct a parallel analysis over the same data with simulated signals added 
and repeat the same process as described above.

The remainder of this section describes key elements of our methods in more
detail.

\subsection{Data}
We assume a linear model for the gravitational wave strain
data~\cite{Finn:1992xs}, \data, which is a vector of discretely sampled time
points for a gravitational wave detector, $j$,
\begin{align}
\data_j = \signal(\params_j) + \noise_j,
\end{align}
where: $\signal(\params_j)$ is an unknown gravitational waveform  accurately
modeled as a function of $\params_j = \{m_1, m_2, a_{1z}, a_{2z}, t_j, \phi_j,
A_j\}$\footnote{These parameters are adequate to describe the measurable
gravitational wave parameters for a non-precessing, circular binary black hole
system with only 2-2 mode emission in a single gravitational wave detector}
with $m_1,m_2$ being the component masses, $a_{1z}, a_{2z}$ being the
orbital-angular-momentum-aligned component spins and $t_j, \phi_j, A_j$ being
the time of coalescence, phase of coalescence, and amplitude, all of which
depend on exactly where the binary is with respect to the $j$th gravitational
wave detector. $\noise_j$ is a realization of detector noise.  As a concrete
example, in this work each job analyzes 800s stretches of data, divided into 4s windows sampled at 2048
Hz. Thus, after including the Fourier transform block length (124s), the dimension of each vector in the work
described in this manuscript is 262144 sample points. In addition, each job also contains start padding (128s),
and stop padding (32s). The templates have at least 6s of zero padding, which makes their length no more than 122s.

We assume that the noise samples are entirely uncorrelated between the
gravitational wave observatories, but that the signals are correlated between
observatories.  In fact, we make the simplifying assumption that the
gravitational waveform is identical between detectors except for an overall
amplitude, $A_j$, time shift, $\Delta t_j$, and phase shift, $\Delta \phi_j$,
~\cite{Allen:2005fk},
\begin{align}
\signal(\params_j) &= \Re\left( \ifft{A_j e^{2 \pi i f \Delta t_j + \Delta \phi_j} \fft{\signal(\inparams,t,\phi)}}\right),
\end{align}
where $\fft{\dots}$ denotes the unitary Fourier transform, and $\inparams{} =
\{m_1, m_2, a_{1z}, a_{2z} \}$. 

The exact realization of noise, $\noise$, is not possible to predict, but we
will assume it is well characterized as a multivariate normal distribution with
a diagonal covariance matrix in the frequency domain, i.e., that it is
stationary.
However later on, particularly in Sec.~\ref{ss:glitch}, we account for the fact
that the data is often not stationary.

\subsection{Spectrum estimation and whitening}

We rely on the same spectrum estimation methods as described
in~\cite{Messick:2016aqy}. Namely we use a median-mean, stream-based spectrum
estimation technique that adjusts to changes in the noise spectrum on
$\mathcal{O(\textrm{min})}$ time scales. The data are divided into 8s blocks
with 6s overlap and the spectrum, $\mathbf{S}_n$ is estimated by windowing the
input blocks with 2s of zero-padding on each side of the window.  Since we
analyze only 800s of data per job, we use a fixed spectrum over the
job duration.

From here forward, we will work in a whitened basis for the data, namely that
\begin{align}
\data_j &\to \fft{\data} \circ \left(\mathbf{S}_n\right)^{-1/2},
\label{eq:whitening}
\end{align}
which implies that all components of \data\ are transformed by the inverse
noise amplitude spectrum.  Therefore, if the amplitude of \signal\ is zero,
\data\ has components that satisfy $p(s_i) = \left(2\pi\right)^{-1/2}\,
e^{-s_i^2/2}$ with $\langle s_i, s_j\rangle = \delta_{ij}$.  In this whitened
data basis, an inner product between two vectors is the dot product $\mathbf{u}
\cdot \mathbf{v}$, and unit vectors are denoted as $\hat{\mathbf{u}}$. We adopt
a normalization such that $\signal \cdot \signal = 1$ and $\langle \noise \cdot
\noise \rangle = \dim{\noise}$.  With these choices the 
SNR is given by $\rho(\params_j) = \signal(\params_j)\cdot \data_j$. We can
evaluate the SNR for the unknown phase and time of coalescence by defining a
complex SNR
\begin{align}
\rho(\inparams, t_j, \phi_j) &= \ifft{\signal(\inparams) \cdot \data_j} \nonumber \\
                           &+ i \ifft{\signal(\inparams, \pi / 2) \cdot \data_j},
\end{align}
which is a \textit{valid} matched-filter output for a duration of time equal to the
length of the data minus the length of the template. With at least 6s of zero-padding, 
the template length is 122s, and with each window using 128s of data, the matched-filter
output is valid for a duration of 6s.

\subsection{Simulation capabilities}

We use the GstLAL data source module~\cite{gstlal}, which provides an interface
into the LAL Simulation package~\cite{lalsuite}.  By providing a LIGO-LW XML
format document containing simulation parameters, we can inject simulated
strain into each of the currently operating ground-based gravitational wave
detectors, LIGO, Virgo, KAGRA and GEO-600.  When operating the pipeline in a
simulation mode, gravitational wave events are reconstructed around a $\pm 2$s
interval around the GPS second of the geocentric arrival time of the
gravitational wave peak strain.

\subsection{Parameter space sampling}

The gravitational wave parameter space is explored
stochastically, with Gaussian jump proposals and refinement steps that
gradually reduce the jump size as the peak in SNR is identified. We will refer to 
this procedure as ``sampling". Our proposal
distribution has a covariance matrix that depends on the location in the
parameter space and the refinement level.  It relies on computing the parameter
space metric, $\g$~\cite{PhysRevD.53.6749}, which is described more in the next
section. We define a sequence of two parameters that control how the sampling
is done, namely $\sigma_k$, which controls the jump size and $N_k$ which
controls the number of samples to reject at each level, $k$, before moving on
to the next.  How exactly to
define these parameters is certainly a topic for future research.  Our choices
here were determined empirically for the particular search we have done.  We
define,
\begin{align}
\sigma_k &= 10^{1-k},\\
N_k &=
\begin{cases}
500 & (k=0) \\
100 & (k>0)
\end{cases},
\end{align}
for $k = 0 \dots n$ where $k$ is terminated based on the mismatch as in step 7 below.
We define the characteristic jump proposal distance as,
\begin{align}
\delta_k(\inparams) := \sigma_k |\g(\inparams)|^{1/8}, 
\end{align}
where $\inparams$ is the set of intrinsic parameters as defined before, and $\delta^{2}_{k}$ is the template
mismatch.

Gravitational waves are searched over 4~s \textit{windows}
of coalescence time using the following procedure. 
\begin{enumerate}
%
\item{
Establish a bounding box in the physical parameter space to search over
} \label{boundarybox}
\item{
Pick a starting parameter point somewhere in the middle of the parameter space. We use the approximate expression for template count in~\cite{PhysRevD.53.6749} to estimate a good central point. 
} 
%
\item{
Evaluate the SNR at this point and set a counter to zero.
}
\item{
Sample from a sampling
function $\Theta(\delta_k, \inparams)$, which is described in detail below in
Sec.~\ref{ss:random}.  \label{randomsample}
}
%
\item{
Check that the new point lies within boundaries established in step~\ref{boundarybox}
and apply any constraint functions. If the point fails to fall within the
constraints, go to step~\ref{randomsample}.
}
%
\item{
Evaluate the SNR at the new point. If the point has a higher SNR than
the previous sample, update the sample and reset the counter to zero.
If the point has a lower SNR, increment the counter.
}
%
\item{
If the counter is less than $N_k$, go back to step~\ref{randomsample}.
If the counter is greater than or equal to $N_k$, check $\delta^2$, where 
$\delta^2$ is the template mismatch between the current and previous sample point.
If $\delta^2 < 0.1$, terminate the sampling. Otherwise, increment $k$, 
reset the counter, and proceed to step~\ref{randomsample}. 
}
\end{enumerate}

\subsubsection{Computation of the binary parameter space metric}
\label{ss:computation}
We define the match between adjacent compact binary waveforms in the space of intrinsic parameters as:
\begin{align}
m(\x, \x + \dx) &= \max_{\phi_c, t_c, A} \left[\, \hat{\signal}(\x) \, \cdot \, \hat{\signal}(\x+\dx) \, \right].
\end{align}
where the maximum is over extrinsic parameters $\{t_c, \phi_c, A\}$. 
Note that $m(\x, \x) = 1$. We also introduce a shorthand for computing the
match along a deviation in only one coordinate as:
\begin{align}
m(\x, \x + \dx_{i}) &= \max_{\phi_c, t_c, A} \left[\, \hat{\signal}(\x) \, \cdot \, \hat{\signal}(\x + \dx_i) \, \right],
\end{align}
where it is assumed that $\Delta\x_{i}$ is nonzero only along a given coordinate
direction.

It has previously been shown~\cite{PhysRevD.53.6749} that is possible to derive a metric on the space of intrinsic
parameters describing compact binary waveforms by expanding our definition of the match locally e.g. about $\Delta \lambda = 0$ as follows,
\begin{align}
&m(\x, \x + \dx) \approx 1 + \nonumber \\
	&\frac{1}{2} \frac{\partial^2}{\partial \Delta\lambda_{i} \partial \Delta\lambda_{j}} m(\x, \x+\dx) \big|_{\dx=0} \Delta\lambda_{i} \Delta\lambda_{j}
\end{align}
which suggests the metric
\begin{align}
g_{ij}(\x) = -\frac{1}{2} \frac{\partial^2}{\partial \Delta\lambda_{i} \partial \Delta\lambda_{j}} m(\x, \x+\dx) \big|_{\dx=0},
\label{eq:metric_coeffs}
\end{align}
The mismatch between templates, $\delta^2 = 1 - m$ becomes
\begin{align}
\delta(\x, \dx)^2 &\approx \dx^T \, \g(\x)  \, \dx,
	\label{eq:mismatch_metric}
\end{align}
In this work, the components of the metric are evaluated with second-order finite differencing,
\begin{align}
g_{ii}(\x) &= -\frac{1}{2} \left[ \frac{m(\x, \x + \Delta \x_i) + m(\x, \x -\Delta \x_i) - 2}{|\Delta \x_i|^2} \right],
\end{align}
and 
\begin{align}
g_{ij}(\x) &= -\frac{1}{2} \frac{1}{4 |\Delta \x_i| |\Delta \x_j|} \times \nonumber \\
	& \biggr[\hspace{6pt} m(\x, \x + \Delta\x_i+\Delta\x_j) \nonumber \\
	& \hspace{10pt} - m(\x, \x + \Delta\x_i-\Delta\x_j) \nonumber \\
	& \hspace{10pt} - m(\x, \x -\Delta\x_i+\Delta\x_j) \nonumber \\
	& \hspace{10pt} + m(\x, \x -\Delta\x_i-\Delta\x_j) \hspace{6pt} \biggr],
\end{align}
for the off diagonal terms. However, we use a more efficient formula for the off diagonal terms,
in which the number of template evaluations is the same, but the number of match calculations is reduced:
\begin{align}
g_{ij}(\x) &= -\frac{1}{2} \frac{1}{2 |\Delta \x_i| |\Delta \x_j|} \times \nonumber \\
        & \biggr[\hspace{6pt} m(\x, \x + \Delta\x_i+\Delta\x_j) - m(\x, \x + \Delta\x_i) \nonumber \\
	& \hspace{10pt} - m(\x, \x + \Delta\x_j) \nonumber +2 - m(\x, \x - \Delta\x_i)\\
        & \hspace{10pt} - m(\x, \x - \Delta\x_j) + m(\x, \x -\Delta\x_i-\Delta\x_j) \hspace{6pt} \biggr]
\end{align}

The sampling method described in section 4 below will not make jumps in coalescence time,
therefore the time component is projected out ~\cite{PhysRevD.53.6749},
\begin{align}
g_{ij}(\x) &\to g_{ij}(\x) - \frac{ g_{ti}(\x) \, g_{tj}(\x) }{ g_{tt}(\x) }.
\end{align}

\subsubsection{Choice of coordinates}

We sought out a coordinate system that maps the masses and spins to be in the
interval $\left[-\infty, \infty\right]$. We also want to choose well measured
physical parameters for mass and spin in at least one dimension each.
Therefore, we use the following coordinates to evaluate the metric
\begin{align}
\lambda_1 &= \log_{10}\left[\frac{(m_1 m_2)^{3/5}}{(m_1 + m_2)^{1/5}}\right] \\
\lambda_2 &= \log_{10}(m_2) \\
\lambda_3 &= \tan\left[\left(\frac{\pi}{2}\right) \left(\frac{a_{1z} m_1 + a_{2z} m_2}{m_1 + m_2}\right)\right] \\
\lambda_4 &= \tan\left[\left(\frac{\pi}{2}\right) \left(\frac{a_{1z} m_1 - a_{2z} m_2}{m_1 + m_2}\right)\right]
\end{align}

\subsubsection{Pathologies of the numerical metric}
\label{ss:eigen}

For certain regions of the parameter space the metric is nearly singular which
leads to numerical errors causing a non positive definite matrix. To fix this, we
conduct an eigenvalue decomposition of $g_{ij}$ 
\begin{align}
g_{ij} &= q_{ik} \, \beta^{k} \, q^{-1}_{kj}
\end{align}
We then define a new set of eigenvalues
\begin{align}
\beta_{\mathrm{min}} &\equiv \max_{k} \left[ \beta^{k} \right] \times \epsilon \\
(\beta^{k})' &=
	\begin{cases}
	\beta_{\mathrm{min}}, & \beta^{k} < \beta_{min} \\
	\beta^{k}, & \text{otherwise},
	\end{cases}
\end{align}
where $\epsilon$ is a parameter which we will call the aspect ratio.
We define the new metric as
\begin{align}
g_{ij} &\to q_{ik} \, (\beta^{k})' \, q^{-1}_{kj}
\end{align}
In practice we find that sampling is better when we artificially distort the
metric by setting $\epsilon = .1$ for the broadest refinement level, and 
$\epsilon = .0001$ for all other levels, and we have done so in this work, though
this should be a direction of future work.

%
%
%


\subsubsection{Drawing random samples from $\Theta(\delta_k, \inparams)$}
\label{ss:random}
When sampling, we desire to have a jump proposal distribution that effectively
probes the space by not making jumps that are either too near or too far.  The
calculation of the parameter space metric $\g$ enables that.  We wish to
propose a jump from $\xs \to \xs + \dxs$ such that the expected mismatch is
$\delta^2$. The metric described in previous sections only applies to the
intrinsic parameters.  For the extrinsic parameters, our jump proposal will
always choose those values of $t$ and $\phi$ which maximize the SNR.
At every accepted jump point, the metric is calculated locally, which requires
21 template evaluations, including the diagonal and off diagonal terms, as
specified in Sec.~\ref{ss:computation}. However, we can afford to calculate
coarse versions of the template waveform, since the match we need to calculate
is between two adjacent templates. This means the waveform calculation cost is not high.
The distance between adjacent templates to calculate the match at, $\Delta\x_{i}$ as
defined in Sec.~\ref{ss:computation} is hardcoded, and is the same for all iterations
of the sampling procedure.

To facilitate jumping in
the intrinsic parameters, we make a coordinate basis transformation in which
the new basis has a Euclidian metric.  The transformation matrix $\M$ will then
be used to transform the coordinates
\begin{align}
\x' &= \M \x.
\end{align}
To solve for $\mathbf{M}$ we rely on the fact that distance is invariant giving
\begin{align}
\delta^2 &= \dx^T \, \g \, \dx \\
         &= (\dx')^T \, \g' \, \dx' \\
         &= (\dx')^T \, \M^T \g \, \M \, \dx'
\end{align}
Setting $\g' = \I$ gives
\begin{align}
\I &= \M^T \g \, \M \\
\M^{-1} (\M^T)^{-1} &= \g \\
\g^{-1} &= \M^T \M
\end{align}
The last line implies that we can solve for \M\ by taking the Cholesky
decomposition of the inverse metric tensor.  Once obtaining \M\ we can produce 
random samples with an expected mismatch by defining,
\begin{align}
\Theta(\delta_k, \inparams) &:= \delta_k, \inparams \to \inparams + \delta_k \M^T \, \vec{\mathcal{N}},
\end{align}
where $\vec{\mathcal{N}}$ is a 4-dimensional vector with random components
satisfying $p(\mathcal{N}_i) = \sqrt{1 / 2 \pi} \exp\left[
-\mathcal{N}_i^2 / 2 \right ]$

\subsubsection{Parameter space constraints}
The previously defined sampling function can produce samples that, while
physical, may be outside of the desired search range.  We implement a series of
user-defined constraints that will reject samples drawn from $\Theta(\delta_k,
\inparams)$.  These are:
\begin{description}
\item[$\mathbf{m_1,m_2,a_1,a_2}$] The user can specify a bounding-box in
component masses and $z$-component spins. Samples outside this bounding box are
rejected.
\item[$\boldsymbol{\eta}$] The user can specify a minimum symmetric mass ratio,
$\eta \equiv (m_1 m_2) / (m_1 + m_2)^2$, below which samples will be rejected.
\item[$\mathcal{M}$] The user can specify a chirp mass range outside of which
samples will be rejected.
\end{description}

\subsubsection{Glitch Rejection}
\label{ss:glitch}
\begin{figure*}[]
\includegraphics[width=\textwidth]{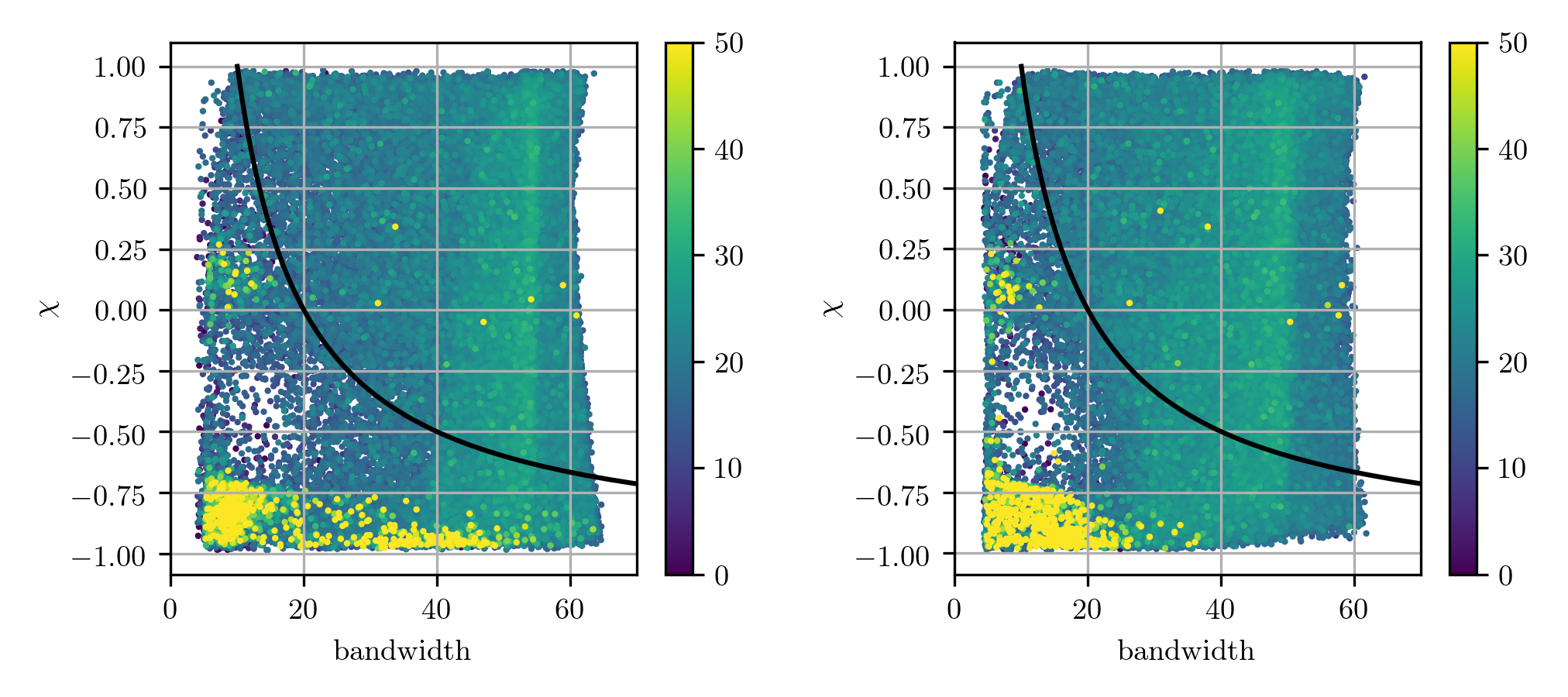}
\caption{\label{fig:glitch_rej}
Triggers found in one month's data for Hanford (left) and Livingston (right), colored by their log-likelihood 
ratio. All the bright points to the right of the boundary are known gravitational wave candidates, 
and all those to the left of the boundary are glitches, and so not considered gravitational wave 
candidates, and not used for background estimation.
}
\end{figure*}

Glitches~\cite{ashton2021parameterised, 2020} are  non-stationarity and non-Gaussian
transient noise artefacts of instrumental or environmental origin found in the data.
We employ a novel data-driven technique to reject short-duration glitches, using two
parameters, the bandwidth, and the effective spin parameter $\chi$.
The bandwidth is the standard deviation of frequency weighted by template amplitude.
It is defined as~\cite{fairhurst2009triangulation}:
\begin{align}
\text{bandwidth}^2 = \frac {\int |\hat{\signal}(\x)|^{2}f^{2}df/\mathbf{S}_n} {\int |\hat{\signal}(\x)|^{2}df / \mathbf{S}_n}
- \left(\frac {\int |\hat{\signal}(\x)|^{2}fdf / \mathbf{S}_n} {\int |\hat{\signal}(\x)|^{2}df / \mathbf{S}_n} \right)^2
\end{align}
whereas $\chi$ is defined as:
\begin{align}
\chi = \frac {m_{1}a_{1z} + m_{2}a_{2z}} {m_{1} + m_{2}}
\end{align}
It has been found that short-duration glitches ring up templates which exclusively
occupy the low bandwidth-low $\chi$ region in bandwidth-$\chi$ space, and that this region
is not occupied by gravitational wave signals. This is illustrated in Fig.~\ref{fig:glitch_rej}. 
As part of the simulation campaign we performed (Refer to Sec.~\ref{ss:simulation} for details),
we found that only 28 injections out of
112526 fell into the glitch region. Minimizing this number by fine-tuning the boundary of the glitch region 
would be a direction for future work.
We define the glitch region as:
\begin{align}
\text{bandwidth} \times (1 + \chi) \leq 20
\end{align}
Any trigger which
falls in this region is not considered as a gravitational wave candidate. Similarly, any time-slid
background samples falling in the glitch region are eliminated, and not used for background estimation.
Triggers are explained in more detail in the next subsection, whereas background estimation 
is explained in section E.

\subsubsection{Computing the log-likelihood ratio, $\mathcal{L}$}

We generally follow the same procedure for ranking candidates as described
in~\cite{Cannon:2012zt, Cannon:2015gha, Hanna:2019ezx} with a couple of notable
exceptions.  First, we only implement a subset of the terms used in the
GstLAL-inspiral pipeline -- it will be the subject of future work to include
more.  Second, we approximate some of the data driven noise terms with analytic
functions.  Third, we adopt a normalization so that for signals, the log
likelihood ratio, $\mathcal{L}$ is approximately $\rho^2 / 2$, where $\rho$ is the network
matched filter SNR defined as the squareroot of the sum of the squares of the
SNRs found in each observatory.  We use the following terms in the log
likelihood ratio:
\begin{description}
\item[$\mathcal{L}(\vec{\rho}, \vec{\xi^2})$] We approximate this term of the
log likelihood ratio as
\begin{align}
\mathcal{L}(\vec{\rho}, \vec{\xi^2}) &= \sum_i \mathcal{L}_i(\rho_i, \xi^2_i) \\
\text{with} \hspace{3pt} \mathcal{L}_i(\rho_i, \xi^2_i) &= \rho_i^2 e^{-4x_i^2 } / 2 -4x_i^2
\end{align}
where $x_i \equiv \max \{0, \xi_i^2 - 1 - 0.0005 \rho_i^2\}$ for each of the $i$
detectors which are assumed to be independent.
The $-4x_i^2$ term acts as a penalty for high $\xi^2_i$ values, and helps eliminate glitches.
\item[$\mathcal{L}(\Delta\vec{t}, \Delta\vec{\phi}, \Delta\vec{\rho})$] for
this term we follow the procedure in~\cite{Hanna:2019ezx} with two
changes. We do not include the $\rho^{-4}$ term. We do this because we are not
constructing a data driven noise term like GstLAL-inspiral, so it's not necessary
to have the corresponding signal term. We also normalize the result to
be 0 when only one detector is operating. This is useful for achieving the
normalization discussed above. These changes have the effect of making
this term $\sim 0$ for things that are consistent with signals. 
\item[$\mathcal{L}(\vec{T})$] this term quantifies the probability of having
``triggered" the combination of the gravitational wave detectors
in which the event was found and is a function of the detectors' sensitivity.  
We will describe triggering in more detail below. For example, it is unlikely 
that only the least sensitive detector would be triggered for a real gravitational
wave event, so this term would be negative in that case. This term is complementary 
to the previous term but accounts for events lacking triggers. 
\item[$\mathcal{L}(\vec{D}_H)$] this term quantifies the relative likelihood
of detecting an event based on the detector horizon BNS distances, $(D_H)_i$.  We
normalize to the horizon distance of LIGO Livingston during O3$ \sim 315$ Mpc.
\begin{align}
\mathcal{L}(\vec{D}_H) &= \ln\left(\frac{\max_i\left[\{(D_H)_i\}\right]}{315 Mpc}\right)^3
\end{align}
%
%
\end{description}
The log-likelihood ratio, $\mathcal{L}$ is then given by
\begin{align}
\mathcal{L} &= \mathcal{L}(\vec{\rho}, \vec{\xi^2}) + \mathcal{L}(\Delta\vec{t}, \Delta\vec{\phi}, \Delta\vec{\rho}) \nonumber \\
            &+ \mathcal{L}(\vec{T}) + \mathcal{L}(\vec{D}_H)
\end{align}

For each sample drawn in step~\ref{randomsample}, we construct a template waveform $\signal(\x)$, and
filter that waveform against the data in each detector stream producing an SNR
time series over a 6s period, including 1s padding on either side.  We then find the peak 
SNR in the middle 4s window in each detector and record the time, phase, SNR, 
and $\xi^2$ of each peak, which we call a ``trigger".  For the collection of triggers, we
cycle through every detector combination - for example, if analyzing \{H,L,V\},
we cycle through \{HLV, HL, HV, LV, H, L, V\} and evalute the likelihood ratio
for each combination.  We then keep the maximum $\mathcal{L}$ found over these detector
combinations. This is done to mitigate the effect of bad data
(noisy data and possibly also glitchy data) in one detector. 
Hence, triggers are obtained by maximizing SNR over 4s windows, whereas the detectors
to be considered for the trigger are obtained by maximizing the likelihood ratio over all possible
detector combinations. Note that the SNR maximization for updating the sample discussed in step 6
is a seperate procedure from either of these.

\subsection{Background estimation}

We treat windows recovered as single triggers and windows recovered as coincidences 
differently while estimating the background. For single trigger windows, the foreground
sample itself is used as the background sample representing that window. To estimate
the coincident background, we form false coincidences from a given
job which analyzes 800s of data in 200 coalescence time
windows. To form false coincidences, we shift the windows in time with respect to
each other. We then draw samples randomly from all
single detector triggers. For each recovered false coincidence, we
compute a $\mathcal{L}$ and histogram the result. This process is then repeated 100
times with different time offsets to increase the amount of background we have.
This background is given an appropriate weight so that the ratio of singles to coincidences
in the background and foreground is the same, as well as to ensure that the background is
normalized. Using the $\mathcal{L}$ histogram for the background,
false alarm rates (FARs) are assigned to all the triggers. One point to note is that
the windows in which we detect events are not used to form combinations so as to not
contaminate the background with signals.

\section{Results}
\subsection{Data set}

We analyze public gravitational wave data from LIGO taken from July 27 00:00
2017 UTC -- Aug 25 22:00 2017 UTC during advanced LIGO's second observing run.
We choose segments of data with a minimum length of 1200s for each of
the LIGO detectors.  From those segments we form coincident segments.  Jobs
require 128s of start padding, 32s of end padding and 124s 
for the Fourier transform block 
to produce triggers. Thus, each job can analyze a minimum of 288s (which produces
triggers for for a single 4s window) and we choose a maximum duration of 1084s 
to produce 800s of triggers over 200 windows. Jobs are overlapped so that
triggers are produced contiguously. 

After accounting for the segment selection
effects, we analyzed approximately 20.17 days of coincident data.

\subsection{Search parameter space}

We search for gravitational wave candidates with component masses between 0.9--400 M$_\odot$ with
$z$-component spins between $-1$ and 1. We conduct the matched filter integration
between 10--1024 Hz.

\subsection{Simulation set}
\label{ss:simulation}

\begin{figure}[h!]
\includegraphics[width=\columnwidth]{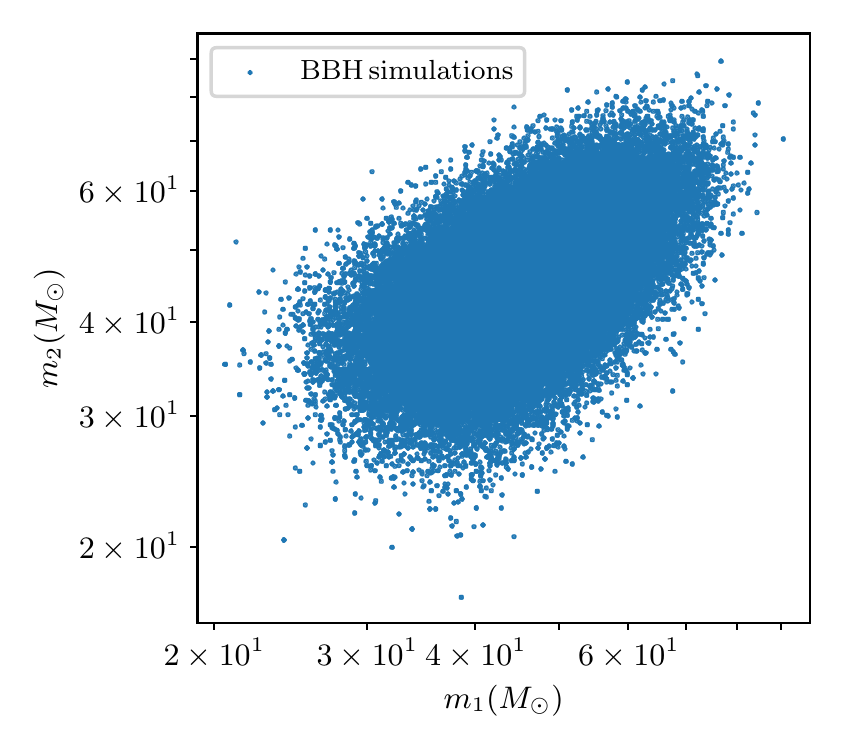}
\caption{\label{fig:inj_masses}
Distribution of component masses as measured at Earth for the BBH
simulation set.  
}
\end{figure}

In order to ascertain our sensitivity to gravitational wave signals of the type
discovered in this data, we conducted a simulation campaign with 112526
simulated signals having a 32 M$_\odot$ mean component mass and standard
deviation of 4.0 M$_\odot$ with aligned dimensionless spins up to 0.25 and a maximum
redshift of 1 isotropically distributed in location. The
injections were distributed uniformly in comoving volume. The red-shifted
component mass distribution is visualized in Fig.~\ref{fig:inj_masses}.
The injection set was specifically created for the BBH parameter space. We do not 
make any claims about the sensitivity of our pipeline 
in other lower mass parameter spaces.

\begin{figure}[h!]
\includegraphics[width=\columnwidth]{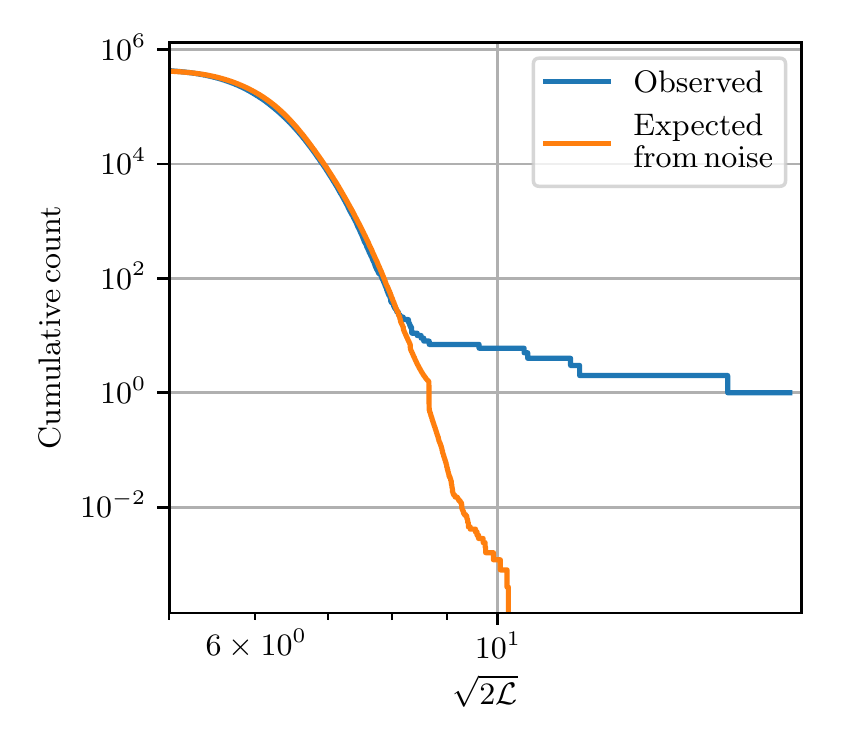}
\caption{\label{fig:candidates}
Cumulative histograms of our search results as a function of likelihood ratio. The orange line represents the corresponding histogram expected from noise during the same time frame.
}
\end{figure}

\begin{figure}[h!]
\includegraphics[width=\columnwidth]{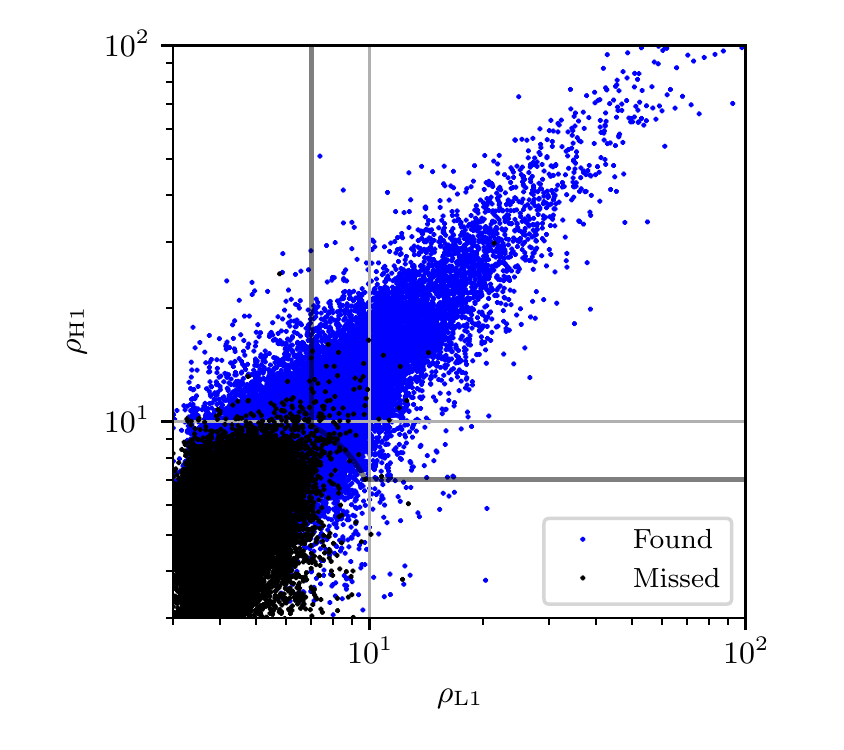}
\caption{\label{fig:inj_dist}
Distribution of injected SNRs for recovered injections above $\mathcal{L} = 35 (\sqrt{2\mathcal{L}} = 8.37$).  Missed injections with network SNR above 12 and detector SNRs greater than 7 (indicated by the shaded contour) are discussed in the appendix.
}
\end{figure}

\subsection{Candidate list}
Our search results are summarized in Fig.~\ref{fig:candidates} and Table~\ref{tab:events}. 
Results from the entire search are shown in Fig. ~\ref{fig:candidates}. In this plot, we show 
the observed distribution of all events as a function of $\sqrt{2 \mathcal{L}}$, an expression 
proportional to the SNR, as well as the background distribution expected from noise during the same time. 
The detected events clearly stand out from the expected noise curve at $\sqrt{2 \mathcal{L}}$ around 8 which 
suggests that the extra events at high $\mathcal{L}$ must be signal-like.

In Table~\ref{tab:events}, we report the ten triggers with the smallest FARs. 
The first five of these events as well as the seventh
were previously reported by the LIGO Collaboration and others ~\cite{LIGOScientific:2018mvr, Nitz:2019hdf, Venumadhav:2019lyq} 
and labelled GW170817, GW170814, GW170809, GW170823, GW170729, and GW170818. These events are detected confidently with FARs of 
$5 \times 10^{-3} yr^{-1}$ for the first five, and $4 \times 10^{-2} yr^{-1}$ for the seventh.
GW170817 is recovered as a single detector candidate in Hanford, since there's a simultaneous glitch in Livingston, and the resulting high $\xi^2$
in Livingston causes its log-likelihood ratio to be strongly penalized.
We report many of the components masses of these events outside of confidence
ranges reported by the LIGO Collaboration ~\cite{LIGOScientific:2018mvr}. It is important to note that this is not a contradiction:
we are not optimizing the posterior probability distribution, as is done during parameter estimation for the results reported by the LIGO Collaboration.
Despite the differences in masses, 
we are able to recover each trigger to within tens of milliseconds of the reported values by the LIGO Collaboration
and are confident they correspond to the respective gravitational wave candidates.

We also recover one binary black hole event, GW170727 previously reported by other groups ~\cite{Venumadhav:2019lyq, Nitz:2019hdf} 
as well as one, GW170817a reported by Zackay et al~\cite{Zackay:2019btq} which do not appear in the LIGO GWTC-2. 
We recover the GW170727 event with a FAR of $3 \times 10^{+1} yr^{-1}$.
We recover GW170817a in Livingston with a FAR of $5 \times 10^{-3} yr^{-1}$
while Zackay reports it with a FAR of $8.7 \times 10^{-2} yr^{-1}$. Zackay also reports the probability of it being of astrophysical origin at 86\% ~\cite{Zackay:2019btq},
but we do not make that estimation here.
As in the previous case, we recover both these events to within tens of milliseconds of the previously reported values 
and are confident that they correspond to the respective gravitational wave candidates.

We make no claims regarding the possibility of the remaining two events we report being gravitational wave candidates. These appear eighth
and ninth in Table~\ref{tab:events}. They are not recovered significantly, and it is likely they are noise.

The first seven events reported in Table~\ref{tab:events}, as well as the last one are excluded from the background, since all of them are previously reported gravitational wave candidates.

\begin{table*}[]
\begin{tabular}{p{0.7in}p{0.4in}p{0.4in}p{0.6in}p{0.6in}p{0.4in}p{0.4in}p{0.4in}p{0.4in}p{0.4in}p{0.4in}p{1.25in}}
FAR (yr$^{-1}$) & $\sqrt{2 \mathcal{L}}$ & $\rho_\text{net}$ & $m_1$ (M$_\odot$) & $m_2$ (M$_\odot$) &  $a_1$ & $a_2$ & $\rho_\text{H1}$ & $\xi^2_\text{H1}$ & $\rho_\text{L1}$ & $\xi^2_\text{L1}$ & Date (UTC) \\
\hline
 $5 \times 10^{-03}$ & 18.5 & 18.7 & 1.8 & 1.1 & $-0.3$ & 0.7 & 18.7 & 0.8 & - & - &2017-08-17 12:41:04 \\
 $5 \times 10^{-03}$ & 16.3 & 17.1 & 38.7 & 24.5 & 0.7 & $-0.9$ & 9.6 & 1.3 & 14.1 & 1.0 &2017-08-14 10:30:43 \\
 $5 \times 10^{-03}$ & 11.9 & 12.6 & 46.4 & 25.8 & 0.6 & $-1.0$ & 6.5 & 1.3 & 10.8 & 0.7 &2017-08-09 08:28:21 \\
 $5 \times 10^{-03}$ & 11.7 & 11.8 & 51.5 & 38.6 & 0.4 & $-0.5$ & 6.6 & 0.9 & 9.8 & 0.7 &2017-08-23 13:13:58 \\
 $5 \times 10^{-03}$ & 10.7 & 10.9 & 73.1 & 43.3 & $-0.1$ & 1.0 & 7.9 & 1.1 & 7.5 & 1.0 &2017-07-29 18:56:29 \\
 $5 \times 10^{-03}$ & 10.6 & 10.7 & 122.7 & 45.5 & 0.9 & $-0.9$ & - & - & 10.7 & 1.0 &2017-08-17 03:02:46 \\
 $4 \times 10^{-02}$ & 9.6 & 10.1 & 39.7 & 36.3 & 0.7 & $-0.8$ & - & - & 10.1 & 1.2 &2017-08-18 02:25:09 \\
 $6 \times 10^{+00}$ & 8.7 & 9.0 & 20.8 & 3.3 & 0.1 & 0.6 & 9.0 & 1.0 & - & - &2017-08-03 05:59:03 \\
 $3 \times 10^{+01}$ & 8.6 & 8.8 & 53.1 & 1.2 & $-0.3$ & 1.0 & 3.7 & 0.9 & 8.0 & 0.8 &2017-08-14 07:35:04 \\
 $3 \times 10^{+01}$ & 8.5 & 8.7 & 51.5 & 43.6 & 0.4 & $-0.9$ & 4.6 & 0.8 & 7.4 & 1.1 &2017-07-27 01:04:30 \\
\end{tabular}

\caption{\label{tab:events}Candidate gravitational wave events with the 10 smallest false alarm rates and largest SNRs. The first seven triggers as well as the last one correspond to known gravitational wave candidates: GW170817, GW170814, GW170809, GW170823, GW170729, GW170817a, GW170818, and GW170727. The other two triggers have not been previously reported as gravitational wave candidates.}
\end{table*}

\begin{figure}[h!]
\includegraphics[width=\columnwidth]{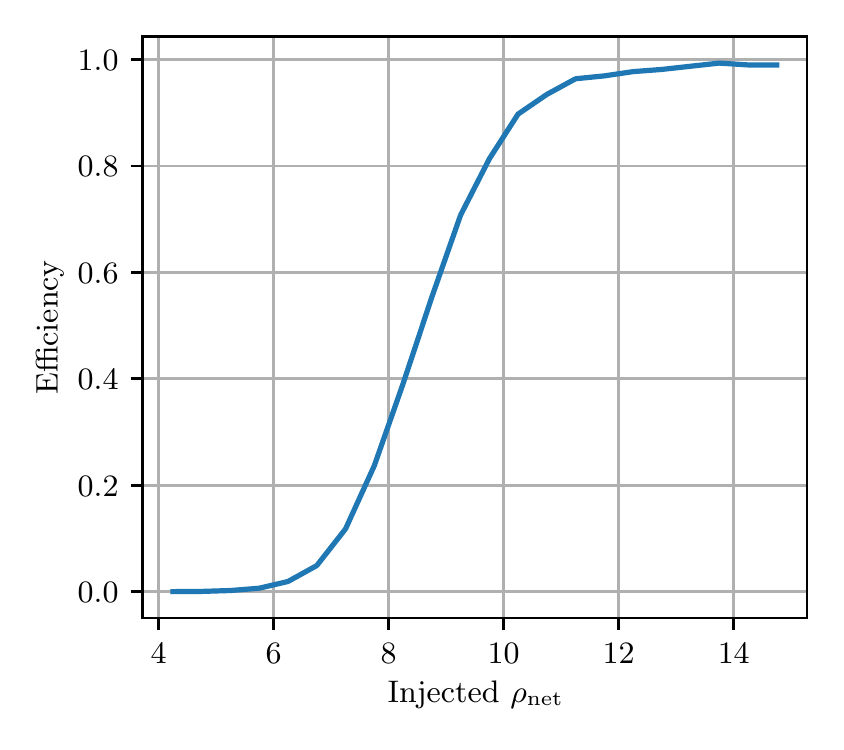}
\caption{\label{fig:eff}
Efficiency of recovering injections at different injected SNRs
}
\end{figure}

\subsection{Sensitivity estimate}

The sensitivity of our new pipeline is demonstrated in Fig.~\ref{fig:inj_dist} and Fig.~\ref{fig:eff}. Fig.~\ref{fig:inj_dist} shows the distribution of all the injected events by SNR with a network SNR of 12 contour, and detector SNR of 7 contours added. This figure shows that the majority of loud injected events were recovered by our pipeline, with 70 missed in the region with network SNR above 12 and detector SNRs greater than 7. Only nine of these missed injections are because the pipeline could not adequately recover the injections. This shows that the pipeline only very rarely gets stuck at local peaks, instead of finding the global maxima, which will correspond to the injected signal. It is possible that as we move to a lower mass parameter space, the frequecy of such occurrences will increase. All of the loud missed events are discussed in more detail in Appendix B.

Fig.~\ref{fig:eff} shows the efficiency of the pipeline as a function of the injected network snr of the synthetic gravitational wave set described in section C. This plot shows that without any data cleaning implementation, almost 90\% of events at SNR 10 are recovered by the pipeline while that percentage only increases with the SNR and plateaus just short of 100\% around SNR 13.

\section{Conclusion}
In this paper, we have described in detail a novel gravitational wave detection algorithm. 
This algorithm searches stochastically over the chosen parameter space, saving the time and computing power required to generate large banks of template waveforms. 
The algorithm samples the parameter space by making jumps with a pre-estimated mismatch between templates informed by the parameter space metric and keeping those points which have a higher SNR. 
This method is shown to be of comparable accuracy in the recovery of gravitational wave events at high masses as current template-based pipelines.

To demonstrate the validity of this method, we have presented an analysis of approximately one month of LIGO data from July 27 00:00 2017 UTC -- Aug 25 22:00 2017 UTC exploring the binary black hole parameter space. 
We recovered six known gravitational wave candidate events to within tens of milliseconds of previously reported coalescence times, as well as two gravitational wave candidates previously reported.

Additionally, we conducted an injection campagin of compact binary mergers to prove the sensitivity of the pipeline to binary black hole merger events. 
We recovered almost 90\% of events with SNR 10 and an increasing percentage at higher SNRs that plateaus just below 100\% at SNR 13. The majority of the missing loud injections were due to the presence of glitches near the injected events.

In the future, we plan to extend our method to all regions of the parameter space. We expect that even though the algorithm will scale similarly to any search using template banks at lower mass, it will still retain its other advantages, such as simpler workflow and ease of setup. We plan to make our method competitive with other searches like GstLAL for LIGO's fourth observing run. It remains an open project to get good convergence during the sampling process for all regions of the parameter space.

\section*{Acknowledgements}
This research has made use of data, software and/or web tools obtained from the Gravitational Wave Open Science Center (https://www.gw-openscience.org/ ), a service of LIGO Laboratory, the LIGO Scientific Collaboration and the Virgo Collaboration. LIGO Laboratory and Advanced LIGO are funded by the United States National Science Foundation (NSF) as well as the Science and Technology Facilities Council (STFC) of the United Kingdom, the Max-Planck-Society (MPS), and the State of Niedersachsen/Germany for support of the construction of Advanced LIGO and construction and operation of the GEO600 detector. Additional support for Advanced LIGO was provided by the Australian Research Council. Virgo is funded, through the European Gravitational Observatory (EGO), by the French Centre National de Recherche Scientifique (CNRS), the Italian Istituto Nazionale di Fisica Nucleare (INFN) and the Dutch Nikhef, with contributions by institutions from Belgium, Germany, Greece, Hungary, Ireland, Japan, Monaco, Poland, Portugal, Spain.

This work was supported by National Science Foundation awards OAC-1841480, PHY-2011865, and OAC-2103662. Computations for this research were performed on the Pennsylvania State University's Institute for Computational and Data Sciences gravitational-wave cluster.  CH Acknowledges generous support from the Eberly College of Science, the Department of Physics, the Institute for Gravitation and the Cosmos, the Institute for Computational and Data Sciences, and the Freed Early Career Professorship.

\appendix \label{sec:appendix}

\section{Data release details and code versions}
A tarball containing the source code and data files necessary to reproduce the results and 
plots in this paper can be found at https://dcc.ligo.org/T2100321. Instructions for 
installing the code and for using it to create the plots and results can be found in 
README.md inside the source\_code directory in the tarball.

\begin{figure}[h!]
\includegraphics[width=\columnwidth]{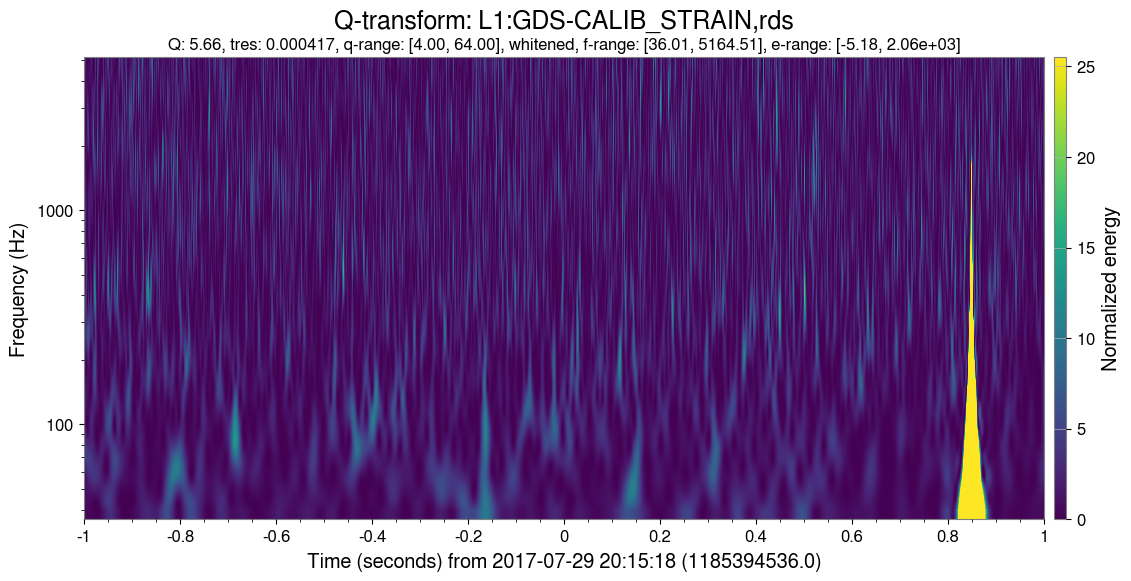}
\caption{\label{fig:inj_glitch}
An example of a Q-transform plot showing a glitch in Livingston, causing a simultaneous injection
to be missed}
\end{figure}

\section{Follow-up of missed injections}
In this appendix, we will discuss the particularly loud injections which were not
recovered during the simulation mode of the pipeline. An injection is deemed to be
recovered, if it was assigned a log-likelihood ratio, $\mathcal{L}$ of 35 or greater.
Out of the 112526 injections, 65361 were missed. Most of these (65291 out of 65361) were missed because
the injected SNR was too low for them to be recovered significantly. Some, however
had a high injected SNR and were still missed. We will discuss the reasons for the same,
for missed injections with network SNR above 12 and detector SNRs greater than 7.
These injections are shown in Fig.~\ref{fig:inj_dist}, of which there are 70. Out of these,
33 were missed due to the data containing a glitch simultaneous to the injection, 
causing the glitch rejection mechanism to reject that part of the data.
The existence of a glitch in the data was verified by creating
Q-transform plots of the data window. An example of such a glitch is shown in Fig.~\ref{fig:inj_glitch}.
Out of the remaining 37 loud missed injections, 28 fell into the glitch region as defined in Section III-D.6,
and hence were rejected. The pipeline failed to recover only 9 injections out of the original 112526.
However, such problematic injections can be recovered by increasing $N_k$, the number of samples to reject 
at each level, $k$, before moving on to the next, at the cost of the run-time of the pipeline.


\bibliography{references}

\end{document}